\documentclass[%
 reprint,
amsmath,amssymb,
aps,
]{revtex4-1}
\usepackage{graphicx}
\usepackage{dcolumn}
\usepackage{bm}
\usepackage{epstopdf}

\begin{document}

\preprint{APS/123-QED}

\title{Structural-transition-induced quasi two-dimensional Fermi surface in FeSe}

\author{Yue Sun,}
 \email{sunyue@issp.u-tokyo.ac.jp}
\author{Tatsuhiro Yamada, Sunseng Pyon, and Tsuyoshi Tamegai}

\affiliation{%
Department of Applied Physics, The University of Tokyo, 7-3-1 Hongo, Bunkyo-ku, Tokyo 113-8656, Japan}
\date{\today}

\begin{abstract}
We report detailed study of angular-dependent magnetoresistance (AMR) with tilting angel $\theta$ from $c$-axis ranging from 0$^\circ$ to 360$^\circ$ on a high-quality FeSe single crystal. A pronounced AMR with twofold symmetry is observed, which is caused by the quasi two-dimensional (2D) Fermi surface. The pronounced AMR is observed only in the orthorhombic phase, indicating that the quasi-2D Fermi surface is induced by the structural transition. Details about the influence of the multiband effect to the AMR are also discussed. Besides, the angular response of a possible Dirac-cone-like band structure is investigated by analyzing the detailed magnetoresistance at different $\theta$. The obtained characteristic field ($B^*$) can be also roughly scaled in the 2D approximation, which indicates that the Dirac-cone-like state is also 2D in nature.

\begin{description}
\item[PACS numbers]
\verb+74.70.Xa+, \verb+74.25.F-+, \verb+72.15.Gd+, \verb+75.47.-m+

\end{description}
\end{abstract}

\pacs{Valid PACS appear here}
\maketitle
\section{introduction}
FeSe has the simplest crystal structure in iron-based superconductors (IBSs), composing of only Fe-Se layers, and shows superconductivity at $\sim$9 K with no need for further doping \cite{HsuFongChiFeSediscovery}. It undergoes only the transition from tetragonal to orthorhombic structure at $T_s$ $\sim$87 K without long-range magnetic order at any temperatures \cite{McQueenPRL}, which is different from the iron pnictides, where the structural transition usually precedes or coincides with antiferromagnetic (AFM) order \cite{FernandesNatPhy}. Such a unique feature makes FeSe an ideal material to study the nematic order, which is often referred as the origin of structural transition and is believed to be related directly to the high-temperature superconductivity \cite{FernandesNatPhy,ChuScience,FradkinAnnu}, without the influence of magnetic order. A splitting of the Fe 3$d_{xz}$ and 3$d_{yz}$ orbitals at the $M$ point of the Brillouin zone is indeed observed by angle-resolved photoemission spectroscopy (ARPES) measurements, and the band splitting is found as large as 50 meV at low temperatures and can persist up to $\sim$110 K above $T_s$, which indicates that the electronic nematicity is caused by the ferro-orbital ordering \cite{NakayamaPRL,ShimojimaPRB}. It is also supported by NMR measurements that spin fluctuations only exist below $T_s$, which is against the spin-driven nematicity \cite{BaekNatMater,BöhmerPRL}.

Besides the simplest structure and the structural transition without magnetic order, which are preferable for probing the mechanism of superconductivity, FeSe attracts much attention also because it provides a promising way to search for superconductors with higher $T_c$. Although the initial $T_c$ in FeSe is below 10 K \cite{HsuFongChiFeSediscovery}, it can be easily increased to 37 K under pressure \cite{MedvedevNatMat} and over 40 K by intercalating spacer layers \cite{BurrardNatMat,SunLilingnature}. Recently, the monolayer of FeSe grown on SrTiO$_3$ is reported to show a sign of superconductivity over 100 K \cite{GeNatMatter}. On the other hand, the Fermi energy $E_F$ of FeSe is found to be extremely small and comparable to the superconducting energy gap $\Delta$, indicating that FeSe is in the crossover region from Bardeen-Cooper-Schrieffer (BCS) to Bose-Einstein-condensation (BEC), which may manifest some unexpected effects \cite{Kasahara18112014}.

To probe those intriguing properties of FeSe, the understanding of its band structure is crucial. ARPES measurements report one small hole pocket at the center ($\Gamma$ point) and one or two electron pockets at the corner ($M$ point) of the Brillouin zone at low temperatures, which is quite different from band structure calculations \cite{NakayamaPRL,MaletzPhysRevB.89.220506,WatsonPRB92,Zhangarvix,SuzukiPhysRevB}. Such a result is also supported by the quantum oscillation measurements \cite{TerashimaPRB}, the mobility spectrum analysis \cite{HuynhPRB} and the three-carrier model fitting to the transport data \cite{WatsonPRL}, although the temperature evolution of the band structure, especially the shrinking and splitting of the electron pocket at $M$ point, is still under debate \cite{Zhangarvix}. More importantly, the ARPES and quantum oscillation results suggest that the electron and hole bands of FeSe at low temperatures may be quasi-2D, which is different from the quasi-3D band structure observed in other IBSs \cite{WatsonPRB92,TerashimaPRB}.

In this report, a pronounced angular-dependent magnetoresistance (AMR) with twofold symmetry is found at low temperatures in FeSe, which proves the quasi-2D nature of the band structure. Temperature evolution of the AMR suggests that the quasi-2D Fermi surface (FS) is induced by the structural transition. Besides, the possible existence of Dirac-cone-like band structure is investigated by measuring the angular dependence of the linear MR, which manifested that the Dirac-cone-like band in FeSe is almost 2D in nature.

\section{experiment}
High-quality FeSe single crystals were grown by the vapor transport method \cite{BöhmerPRB}. The obtained crystals show high-quality with sharp superconducting transition width $\Delta$$T_c$ $<$0.5 K from susceptibility measurements, and large residual resistivity ratio (RRR = $\rho$ (300 K)/$\rho$ (10 K)) $\sim$33 as reported in our previous publications \cite{SunPhysRevBJcFeSe,SunPhysRevB.93.104502}. Transport measurements were performed by using the six-lead method with a physical property measurement system (PPMS, Quantum Design). In order to decrease the contact resistance, we sputtered gold on the contact pads just after the cleavage, then gold wires were attached on the pads with silver paste, producing contacts with ultra-low resistance ($<$100 $\mu\Omega$). Details about the AMR measurements were shown as a sketch in the inset of Fig. 1(a). The crystal was mounted on a rotating stage so that the angle $\theta$ between the $c$-axis of the crystal and magnetic field can be continuously changed from 0$^\circ$ to 360$^\circ$. The excitation current ($I$) flowing in the $ab$-plane was kept always perpendicular to the field. Since the AMR was measured with $\theta$ tilting from the $c$-axis, the twin boundaries in the $ab$-plane affect little.

\section{results and discussion}
\begin{figure}\center
\includegraphics[width=8.5cm]{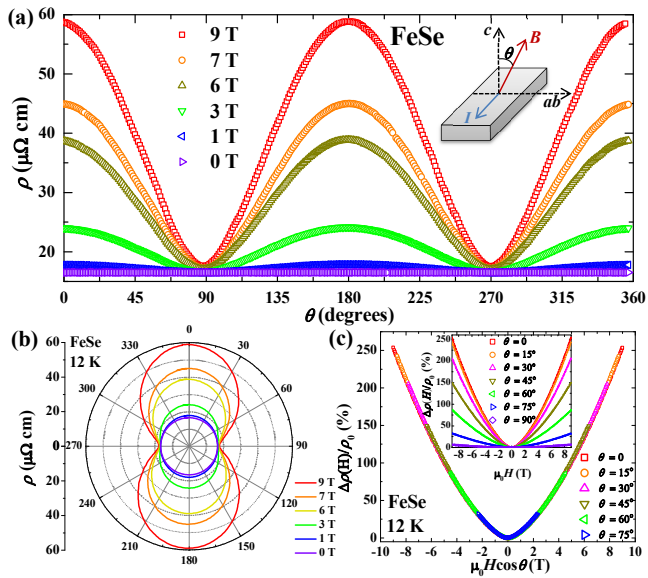}\\
\caption{(Color online) (a) Angular dependent in-plane resistivity $\rho$ of FeSe measured at 12 K under 0, 1, 3, 6, 7, 9 T. The inset shows the configuration of the measurement. (b) Polar plot of the AMR data in (a). (c) The inset is the field dependent MR measured at 12 K with $\theta$ = 0$^\circ$, 15$^\circ$, 30$^\circ$, 45$^\circ$, 60$^\circ$, 75$^\circ$, and 90$^\circ$. Main panel is the 2D scaling of the MR by $\mu_0H$cos$\theta$.}\label{}
\end{figure}

Angular-dependent magnetoresistance for FeSe measured under different magnetic fields at 12 K is shown in Fig. 1(a). Obviously, the MR of FeSe exhibits a significant angular dependence. It reaches the maximum values when the field is parallel to the $c$-axis ($\theta$ = 0$^\circ$ and 180$^\circ$), and gradually decreases with tilting angle $\theta$ reaching the minimum values when the field is perpendicular to the $c$-axis ($\theta$ = 90$^\circ$ and 270$^\circ$), which follows the shape of $|$cos$\theta$$|$ curve. Such behavior can be seen more clearly in the polar plot as shown in Fig. 1(b) that the angular oscillations of MR manifest obvious twofold symmetry, and the anisotropy of MR becomes stronger with increasing the applied magnetic field.

The response of the charge carriers to the applied field, like the magnitude of MR is determined by the component of their mobility in the plane perpendicular to the magnetic field. For materials holding only isotropic three-dimensional (3D) FSs, there should be no obvious angular dependence of MR since the mobility is isotropic. For materials with anisotropic 3D FSs, the AMR reflects the magnitude of the anisotropy, and the FSs topology, which is more complex and has no unified symmetry. For example, the AMR of Bi shows $\pi$/3 periodicity of the angular oscillations when the current applied along the trigonal axis because of the three anisotropic electron bands separated by 2$\pi$/3 with each other in the binary/bisectrix plane \cite{ZhuNatPhyBi}. On the other hand, for the (quasi-)2D FS, the AMR should only respond to the magnetic field component perpendicular to the 2D plane \cite{GrigorievPRB}. When the applied field is perpendicular to the 2D plane, the charge carriers moves in the 2D plane. In this configuration, the Fermi velocity of those charge carriers and the Lorentz force they feel are maximum, which give rise to the largest value of MR. With tilting the direction of the field to the 2D plane, the value of MR will be gradually decreased because the field component perpendicular to the 2D plane is reduced. Thus, the AMR shows twofold symmetry and is proportional to the component of $B$$|$cos$\theta$$|$ \cite{GrigorievPRB}. Such behavior is indeed observed in some quasi-2D systems such as the Sr$_2$RuO$_4$ \cite{OhmichiPRB}, $\alpha$-(BEDT-TTF)$_2$I$_3$ \cite{OhmichiJPSJ}, Sr(Ca)MnBi$_2$ \cite{WangPRB,WangCaMnBi}, LaAgBi$_2$ \cite{WangLaAgBi}, and the surface state of topological insulators \cite{AnalytisNatPhy,QuScience}. Obviously, the AMR results of FeSe shown in Fig. 1(a)-(b) obey the behavior of quasi-2D system, indicating the two-dimensional Fermi surface is dominating the transport properties at low temperatures.

To get more direct and quantitative evidence for the quasi-2D AMR, we also measured the magnetic field dependent MR from -9 T to 9 T with fixed angles $\theta$ = 0$^\circ$, 15$^\circ$, 30$^\circ$, 45$^\circ$, 60$^\circ$, 75$^\circ$, 90$^\circ$ at 12 K. As shown in the inset of Fig. 1(c), MR (= $(\rho(H)-\rho(0))/\rho(0)$) for FeSe reaches a large value over 200\% at 12 K under 9 T when $\theta$ = 0$^\circ$, which is similar to previous reports \cite{HuynhPRB,WatsonPRL}. And the magnitude of MR decreases gradually with increasing $\theta$. As we already explained above and was proved previously in other quasi-2D materials \cite{AnalytisNatPhy,FangLPhysRevB.90.020504}, the AMR originated from quasi-2D FS should be only proportional to the perpendicular component of field. In this case, the MR measured under different angles can be simply scaled by $B$cos$\theta$. To test this assumption, we replotted the data of MR versus $\mu_0H$cos$\theta$ in the main panel of Fig. 1(c). It is obvious that the MR can be well scaled onto a unique curve, which strongly proves the dominance of the quasi-2D FSs at low temperatures in FeSe. This finding is also supported by the recent quantum oscillation, and the ARPES results that the observed FSs of FeSe are quite different from the band structure calculation, consisting of only quasi-2D hole- and electron-type tiny cylinders along $k_z$ direction \cite{TerashimaPRB,WatsonPRB92}. Actually, the multiband nature of FeSe containing both electron- and hole-type pockets complicates the understanding of the AMR results. About this point, we will discuss it in more detail later.

\begin{figure}\center
\includegraphics[width=8.5cm]{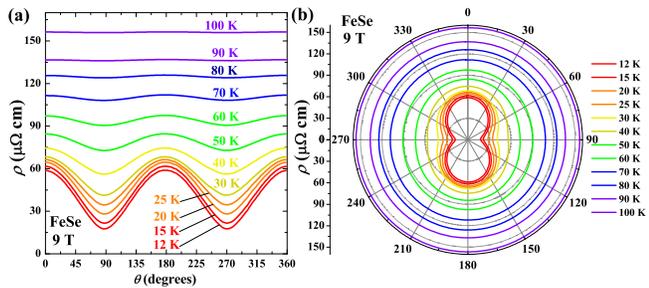}\\
\caption{(Color online) (a) Angular dependent in-plane resistivity $\rho$ of FeSe measured under 9 T at temperatures from 12 $\sim$ 100 K. (b) The polar plot of the AMR data in (a).}\label{}
\end{figure}

In order to get more comprehensive understanding of the quasi-2D FSs in FeSe, we  also measured the temperature evolution of the AMR. Typical results of the AMR at temperatures ranging from 12 K to 100 K are shown in Fig. 2(a), and the corresponding polar plot is shown in Fig. 2(b). The AMR keeps twofold symmetry at low temperatures, and the angular oscillations are gradually smeared out with increasing temperature. When the temperature is increased over 80 K, the oscillation becomes almost negligible.

To observe the temperature evolution of the AMR more clearly, we calculate the differences of the resistivities ($\Delta\rho$) measured at 9 T in $\theta$ = 0$^\circ$ and 90$^\circ$, and depict in the main panel of Fig. 3(a). The $\Delta\rho$ manifests the magnitude of AMR, i.e. the two dimensionality of FS. For comparison, we also show the temperature dependence of resistivity and Hall coefficient in the inset. Details about the transport properties can be found in our previous report \cite{SunPhysRevB.93.104502}. An obvious kink-like behavior related to the structural transition (breaking the tetragonal $C_4$ lattice symmetry down to orthorhombic $C_2$ symmetry) is observed at the temperature of $\sim$86 K similar to the previous report \cite{Kasahara18112014}, and is also marked by the arrow in the main panel of Fig. 3. Above the structure transition temperature $T_s$, $\Delta\rho$ is close to zero. By contrast, the value of $\Delta\rho$ increases steeply with decreasing temperature below $T_s$, i.e. the two dimensionality becomes more dominant. On the other hand, the 2D scaling of MR with different $\theta$ is satisfied at any temperature below $T_s$. Two typical scaling results of data at 12 K and 60 K are shown in Fig. 1(c) and Fig. 3(b), respectively. However, such scaling becomes invalid at temperature above $T_s$, like 100 K, as shown in the Fig. 3(c), which manifests that the FSs of FeSe above $T_s$ is no more quasi-2D.

\begin{figure}\center
\includegraphics[width=8.5cm]{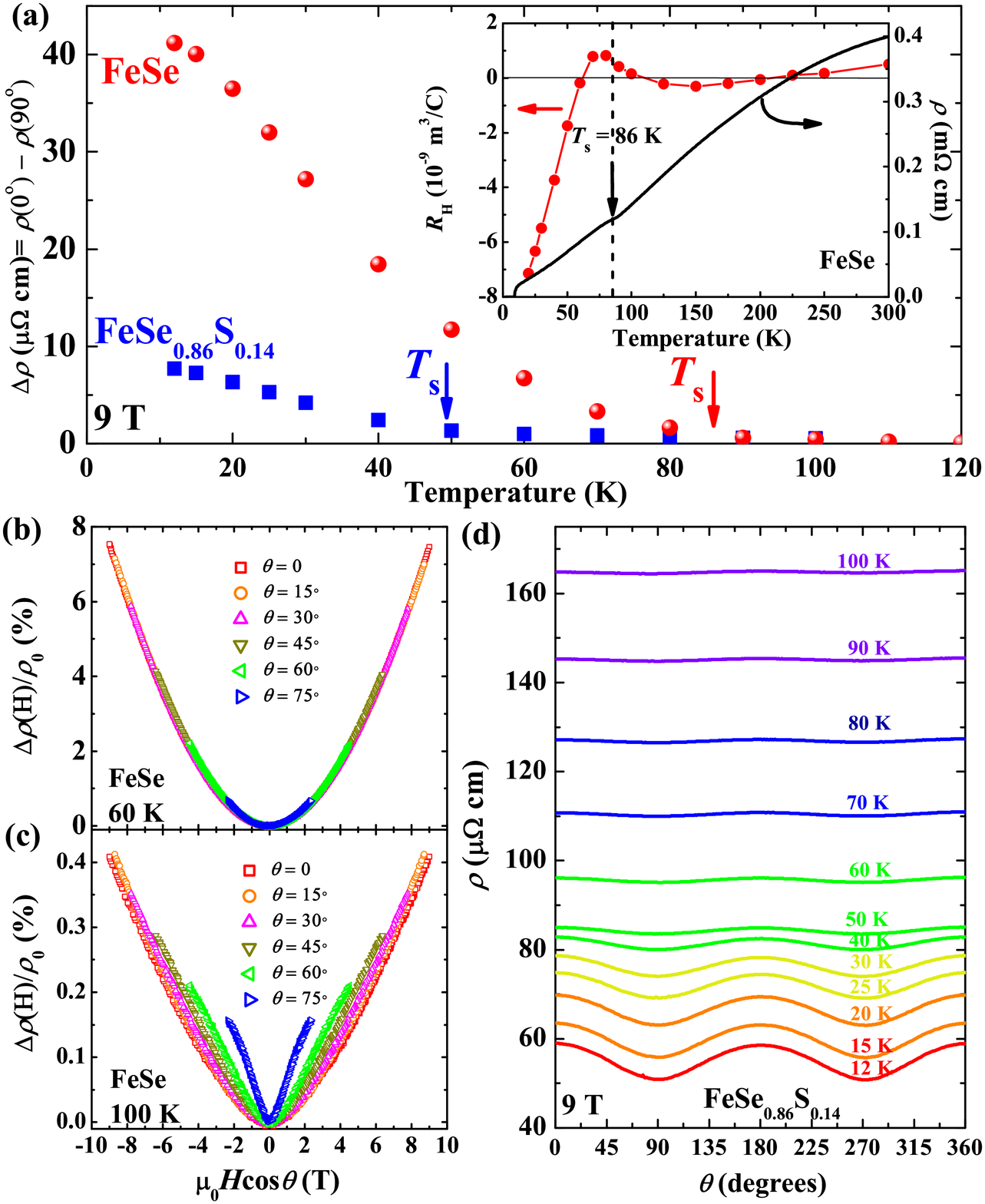}\\
\caption{(Color online) (a) Temperature dependence of the differences of resistivities measured in $\theta$ = 0$^\circ$ and 90$^\circ$ for FeSe and FeSe$_{0.86}$S$_{0.14}$ under 9 T. The inset shows the temperature dependence of the zero-field resistivity and Hall coefficients. (b) and (c) are the 2D scaling of the AMR at 60 K and 100 K by $\mu_0H$cos$\theta$. (d) Angular dependent in-plane resistivity $\rho$ of FeSe$_{0.86}$S$_{0.14}$ measured under 9 T at temperatures from 12 to 100 K.}\label{}
\end{figure}

\begin{figure*}\center
\includegraphics[width=18cm]{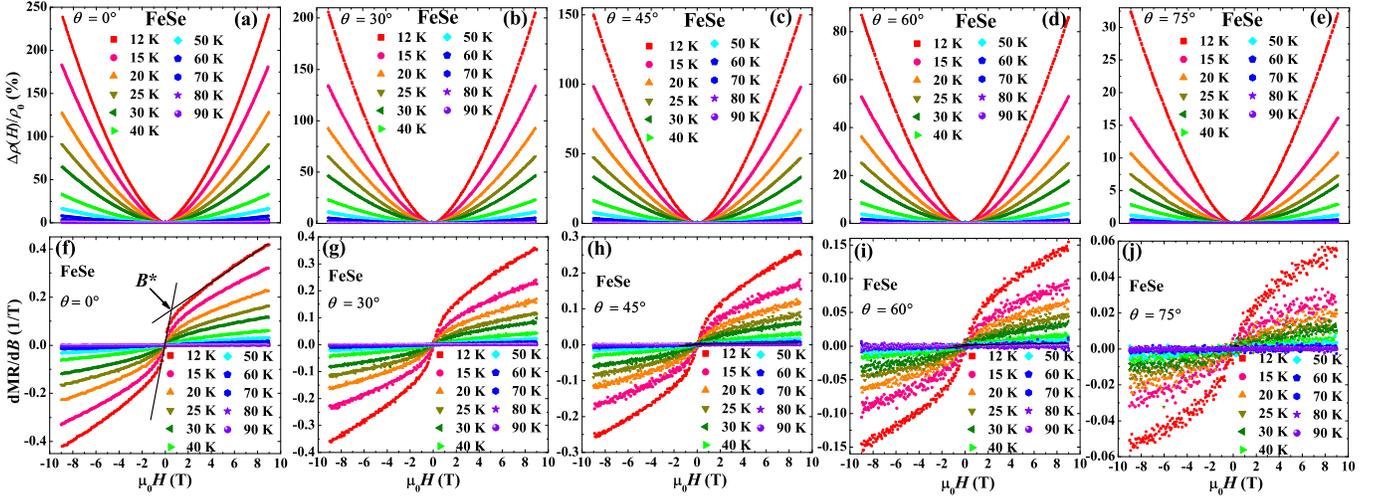}\\
\caption{(Color online) Magnetic field dependence of  MR (= $(\rho(H)-\rho(0))/\rho(0)$) for FeSe with $\theta$ = (a) 0$^\circ$, (b) 30$^\circ$, (c) 45$^\circ$, (d) 60$^\circ$, and (e) 75$^\circ$. The corresponding field derivative MR (= dMR/d$B$) is shown in (f) - (j), respectively.}\label{}
\end{figure*}

The S-doping has been found to suppress the structure transition temperature of FeSe without large modification on the electronic or superconducting properties \cite{AbdelPRB}, which gives us a good opportunity to testify the relation between the quasi-2D FS and the structure transition. Hence, the AMR measurements were also performed on the FeSe$_{0.86}$S$_{0.14}$ single crystal with $T_s$ $\sim$49 K. More information about this crystal can be seen in our previous report \cite{SunPhysRevB.93.104502}. The twofold symmetric AMR is also observed in FeSe$_{0.86}$S$_{0.14}$ at low temperatures as shown in Fig. 3(d). The temperature dependence of the $\Delta\rho$ is plotted in Fig. 3(a). Similar to FeSe, $\Delta\rho$ of FeSe$_{0.86}$S$_{0.14}$ also manifests an obvious increase when the temperature is reduced just below $T_s$. Above $T_s$, $\Delta\rho$ keeps a small value close to zero. All the above results indicate that the emergence of quasi-2D FS is induced by the structure transition. A dramatic splitting of the Fe 3$d_{xz}$ and 3$d_{yz}$ orbitals was found beginning at the temperatures above $T_s$, which is thought as the main driven force of the structure transition \cite{NakayamaPRL,ShimojimaPRB,BöhmerPRL}. Meanwhile, the Fermi surfaces were elongated during the splitting as observed by the ARPES results \cite{WatsonPRB92}. Thus, the quasi-2D FSs in FeSe may come from the band reconstruction induced by the orbital-ordering. Such explanation is also supported by the effect of S doping. As shown in Fig. 3(a), although the $\Delta\rho$ of FeSe$_{0.86}$S$_{0.14}$ shows similar behavior as FeSe, its absolute value at low temperatures is reduced to less than 25\% of that for FeSe, which indicates that the quasi-2D becomes weaker after S doping. Such a result is consistent with the ARPES observation that S doping reduces the Fermi surface anisotropy, and suppresses the orbital ordering \cite{WatsonPRB91}. It should be noted that a possible temperature-induced Lifshitz transition is suggested in FeSe$_{1-x}$S$_x$ ($x$ = 0.055) at a temperature higher than $T_s$ by recent ARPES measurements \cite{Abdel-HafiezPhysRevBImpurity2016}.

Another possible mechanism is the Pomeranchuk instability, which will spontaneously deform the FS along a specific direction, is also proposed as the origin of the nematic transition in FeSe based on the recent electronic Raman scattering measurements \cite{Massatarxiv}. The nematic state, spontaneous symmetry breaking from $C_4$ to $C_2$ symmetry, is one of the key issues of the IBSs since it may be directly related to the mechanism of high $T_c$ superconductivity. However, its origin is still under debate.  In iron-pnictides, most results support that the nematic state have the magnetic origin from the electron's spin \cite{FernandesNatPhy}. On the contrary, the magnetism may not be the main driving force for the nematic phase in FeSe since long-range antiferromagnetic order does not exist at any temperatures, and the magnetic fluctuations are only observed below $T_s$ \cite{NakayamaPRL,ShimojimaPRB,BöhmerPRL}. It most likely to have the orbital origin based on the recent experimental results \cite{BaekNatMater,WatsonPRB92}. Actually, a nematic quantum critical point (QCP) has recently been observed in FeSe$_{1-x}$S$_x$ system when the structural transition is totally suppressed by S doping, while the nematic fluctuations are found to be strongly enhanced \cite{HosoiPNAS}. Furthermore, $T_c$ shows a maximum deep inside the nematic ordered phase rather than near the QCP. Thus, the value of $T_c$ is not simply related to the nematic order or its fluctuations, suggesting that the AFM fluctuations may also contribute to the enhancement of $T_c$ in FeSe$_{1-x}$S$_x$ system. Our AMR results show that not only the rotational symmetry but also the Fermi surface topology is drastically altered by the orbital order, which may be another clue to probe the origin of the nematic state.

Now, we discuss the influence of multiband nature to the AMR in FeSe. As is well known, the FeSe contains at least one hole-type and one or two electron-type bands at temperatures below $T_s$ \cite{WatsonPRB92}. Thus, the AMR reflects the joint contributions from all the different bands. As already shown in Fig. 1(c), the AMR at 12 K can be well scaled by the 2D scaling, which means that all the bands at that temperature should be with the quasi-2D structure, otherwise the scaling will be violated by the contribution with either the isotropic or the anisotropic 3D structures. On the other hand, the Hall effect results shown in the lower inset of Fig. 3 manifests that the electron-type bands are dominant at 12 K. In this case, if only one electron pocket exists at that temperature, the electron pocket should be quasi-2D. If two electron bands exist and are comparable in size, both bands should be quasi-2D in nature. However, recent ARPES results show that the d$_{yz}$ band in the nematic state shifts up around the $M_x$ point, while the d$_{xz}$ band shifts downwards around the $M_y$ point and opens a hybridization gap with the d$_{xy}$ band, which enlarges the electron pocket at the $M_y$ point and largely compresses that at $M_x$ point \cite{Zhangarvix}. Thus, the scaling of AMR at 12 K can only confirm the quasi-2D structure of the electron pocket at $M_y$ point. To understand the structure of other bands, we also check the 2D scaling of MR at different temperatures. Shown in the Fig. 3(b) is the scaling results at 60 K, where the contributions from electron- and hole-type bands are almost equal since the Hall coefficient is close to zero. As we mentioned before, the AMR data at 60 K can be also well scaled similar to the case of 12 K, which means that the hole-type band is also quasi-2D below $T_s$. Since the orbital-ordering-triggered band reconstruction starts at around $T_s$, the sizes of the pockets at $M_x$ and $M_y$ points are comparable at temperatures close to $T_s$, as can be seen in Fig. 2(c) of Ref. \cite{Zhangarvix}. Thus, the possible existed electron bands at $M_x$ point is also quasi-2D in nature. Our observation of the quasi-2D structure for all the different bands of FeSe below $T_s$ is consistent with the recent ARPES and quantum oscillation results \cite{WatsonPRB92,TerashimaPRB}.

One more intriguing feature of FeSe is the possible existence of Dirac-cone-like band dispersion with ultrahigh mobility, which is supported by the ARPES \cite{Zhangarvix}, mobility spectrum analysis  \cite{HuynhPRB}, three-carrier model fitting  \cite{WatsonPRL}, and the recent band structure calculations \cite{Onariarxiv}. For the material with Dirac-cone state, the gap between the zeroth and first Landau levels $\Delta_{LL}$ is described as $\Delta_{LL}$ = $\pm$$v_F$$\sqrt{2e\hbar B}$ \cite{AbrikosovPRB}, leading to a much larger Laudau level (LL) splitting compared with the conventional band structure where $\Delta_{LL}$ = e$\hbar B$/$m^*$. Consequently, the quantum limit where all the carriers occupy only the lowest LL \cite{AbrikosovPRB,AbrikosovEPL} can be achieved in the low field region. In such a case, the MR of the material with Dirac-cone state usually increases linearly with magnetic field as already observed in graphene \cite{NovoselovNature}, topological insulators \cite{TaskinPRL}, surface state of W(110) \cite{MiyamotoPhysRevB.93.161403}, and some layered compounds with two-dimensional Fermi surface (like SrMnBi$_2$) \cite{ParkPRL,WangPRB}. Such a linear MR component has also been observed in FeSe and found to be triggered by the structural transition \cite{SunPhysRevB.93.104502}, which supports the existence of Dirac-cone-like band dispersion.

\begin{figure}\center
\includegraphics[width=8.5cm]{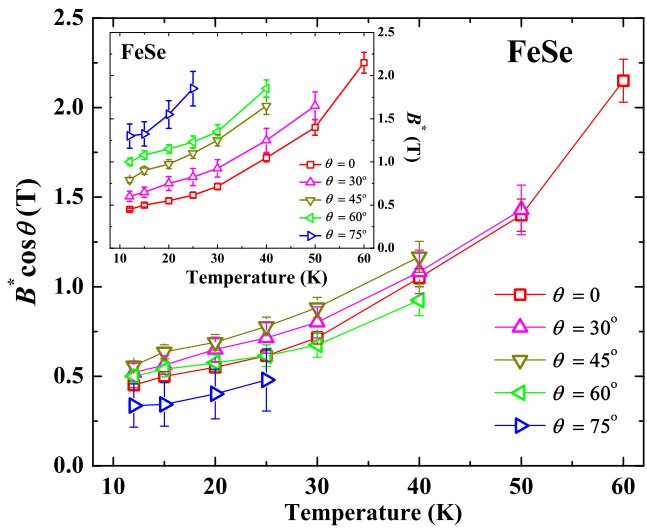}\\
\caption{(Color online) Temperature dependence of the characteristic field $B^*$ (inset), and the scaled $B^*$cos($\theta$) (main panel) for FeSe measured with $\theta$ = 0$^\circ$, 30$^\circ$, 45$^\circ$, 60$^\circ$, and 75$^\circ$.}\label{}
\end{figure}

To know the angular resolution of the possible Dirac-cone-like band structure, we measured the field dependent MR at different temperatures and $\theta$. Fig. 4(a)-(e) shows the MR result of FeSe at temperatures ranging from 12 K to 90 K with $\theta$ = 0$^\circ$, 30$^\circ$, 45$^\circ$, 60$^\circ$, and 75$^\circ$, respectively. Obviously, the MR at temperatures higher than $T_s$ shows a relatively small value, $\leq$ 1\%, while it increases dramatically at temperatures below $T_s$, and reaches a large value, for example over 200\% at 12 K under 9 T when $\theta$ = 0$^\circ$. The large value of MR at low temperatures is similar to previous reports \cite{HuynhPRB,WatsonPRL}. More interestingly, the MR below $T_s$ tends to increase with magnetic field in a more linear fashion at high fields, which can be seen more clearly in the first-order derivative d($MR$)/d$B$ as shown in Fig. 4(f)-(j). As marked by the solid lines in Fig. 4(f), d($MR$)/d$B$ linearly increases with magnetic field at small fields, which indicates a classic $B^2$ dependence of MR. On the other hand, above a characteristic field $B^*$, d($MR$)/d$B$ suddenly saturates to a much reduced slope. As already explained in our previous publication \cite{SunPhysRevB.93.104502}, d($MR$)/d$B$ of FeSe shows a reduced slope above $B^*$ rather than a field-independent plateau due to the multiband structure containing normal band with mobility comparable to the Dirac fermions.  Actually, such behavior of the reduced slope in MR is also observed before in other compounds like Sr(Ca)MnBi$_2$ \cite{WangPRB,WangCaMnBi} and Ba(Sr)Fe$_2$As$_2$\cite{KuoPRB,ChongEPL}. To accurately estimate the values of $B^*$, we fitted the magnetic field dependent MR data by the Eq. (1) of Ref. \cite{SunPhysRevB.93.104502}. The value of $B^*$ can be obtained as the crossing point of the fitting curves for the low field ($\mu_0H <$ 0.5 T) and high field ($\mu_0H >$ 4 T) MR regions.

The temperature dependence of $B^*$ at different $\theta$ for FeSe is shown in the inset of Fig. 5. (Some data at high temperatures with large $\theta$ is not included because the two-slope behavior becomes obscured.) Obviously, $B^*$ gradually increases with increasing of $\theta$, which means that the required magnetic field to split the LL becomes larger when the direction of the field tilts away from $k_z$. To see the relation between the $B^*$ and $\theta$ more clearly, we replot the data as $B^*$cos($\theta$) vs temperature in the main panel of Fig. 5. The $B^*$ with different $\theta$ tends to be scaled into one curve within the extent of deviation. Such a deviation may come from the influence of the in-plane MR ($\theta$ = 90$^\circ$), which is almost negligible in the 2D scaling of AMR because its magnitude is much smaller than that at $\theta$ = 0$^\circ$ as can be seen in the inset of Fig. 1(c). However, it disturbs the determination of $B^*$, especially in the situation of large $\theta$, where the value of MR is relatively smaller.

As proposed by the ARPES and band structure calculation results, the Dirac-cone-like band may come from the band shift, which is caused by ferro-orbital ordering. In detail, the d$_{xz}$ band in the nematic state shifts downwards around the $M_y$ point, and opens a hybridization gap with the d$_{xy}$ band, which enlarges the electron pocket at the $M_y$ point, while the d$_{yz}$ band shifts up around the $M_x$ point, and deforms the electron pocket at $M_x$ point into two Dirac-cone-like pockets \cite{Zhangarvix,Onariarxiv}. Combining the ARPES measurements, band structure calculations, and our observations here, the possible Dirac-cone bands in $M_x$ point may function in two dimensions, i.e. in the $ab$-plane. Since the linear MR is only a indirect evidence of the Dirac-cone state, more direct experiments such as ARPES, especially along $k_z$ direction, is required to clarify this issue.

\section{conclusions}
In summary, we investigated the AMR on high-quality FeSe single crystal with angel $\theta$ tilting from 0$^\circ$ to 360$^\circ$ with $c$-axis. A pronounced AMR with twofold symmetry was observed at low temperatures, and was proved to be originated from the quasi-2D FSs because of the successful 2D scaling of the MR by $\mu_0H$cos$\theta$. Such a pronounced AMR is observed only in the orthorhombic phase, indicating that the quasi-2D FSs in FeSe are induced by the structural transition. Furthermore, the successful 2D scaling of AMR at all temperatures below $T_s$ suggests that both the hole and electron type bands are quasi-2D in nature below $T_s$. Besides, a linear contribution of the field dependent MR is observed at different $\theta$. The obtained characteristic field, $B^*$ can be also roughly scaled by the 2D scaling, which indicates that the possible Dirac-cone state is also 2D in nature.

\acknowledgements
Y.S. gratefully appreciates the support from Japan Society for the Promotion of Science.

\bibliography{references}

\end{document}